\documentclass[preprint,12pt]{elsarticle}

\usepackage{graphicx} 
\usepackage{ascmac,amsmath,amssymb,mathtools,siunitx}
\usepackage{cleveref}
\usepackage{xspace}
\usepackage{natbib}

\newcommand{\aNPD}{$\alpha$-NPD\xspace}
\newcommand{\intcharge}[1]{\SI{#1}{\milli \coulomb \per \square \meter}}
\newcommand{\GSPslope}[1]{\SI{#1}{\milli \volt \per \nano \meter}}

\journal{Organic Electronics}

\begin{document}
 
\begin{frontmatter}

\title{Does Fermi Level Alignment Hold Across Organic Interfaces? — An Investigation Using a Rotary Kelvin Probe}

\author[label1]{Masahiro Ohara}
\ead{moohara@shinshu-u.ac.jp}
\author[label2]{Taiyo Inoue}
\author[label5]{Yuya Tanaka}
\author[label2,label3,label4]{Hisao Ishii}

\affiliation[label1]{organization={Faculty of Engineering, Shinshu University},
            addressline={4-17-1 Wakasato}, 
            city={Nagano-shi},
            postcode={3808553}, 
            state={Nagano},
            country={Japan}}
            
\affiliation[label2]{organization={Graduate School of Science and Engineering, Chiba University},
            addressline={1-33 Yayoi-cho}, 
            city={Inage-ku},
            postcode={2638522}, 
            state={Chiba},
            country={Japan}}

\affiliation[label3]{organization={Center for Frontier Science, Chiba University},
            addressline={1-33 Yayoi-cho}, 
            city={Inage-ku},
            postcode={2638522}, 
            state={Chiba},
            country={Japan}}

\affiliation[label4]{organization={Molecular Chirality Research Center, Chiba University},
            addressline={1-33 Yayoi-cho}, 
            city={Inage-ku},
            postcode={2638522}, 
            state={Chiba},
            country={Japan}}

\affiliation[label5]{organization={Graduate School of Science and Technology, Gunma University},
            addressline={1-5-1 Tenjin-cho}, 
            city={Kiryu-shi},
            postcode={3768515}, 
            state={Gunma},
            country={Japan}}
            
\begin{abstract}
Understanding energy level alignment at organic interfaces is crucial for optimizing the performance of organic devices. Interface dipole and band bending significantly influence carrier recombination and generation mechanisms. 
A method of simulating energy level alignment at metal/organic and organic/organic interfaces by assuming a thermal equilibrium model has been proposed, but its validation against experimental methods is still limited.
In this study, the work function change in the \aNPD/HAT-CN/Au interface was measured as a typical donor/acceptor system using a rotary Kelvin probe (RKP). Our findings demonstrate good agreement with simulations only at metal/organic interfaces which have "active" charge transfer. It is suggested that thermal equilibrium is not achieved simply by depositing the film under dark condition, and some treatment to supply carriers, such as exposure to UV light, is necessary for accurate evaluation. At the organic/organic interface, the the experimental results did not agree with thermal equilibrium model, highlighting the need to consider substrate-driven carrier supply and polarization effects when evaluating energy level alignment.

\end{abstract}


\end{frontmatter}


\section{Introduction}
Precise measurement of energy level alignment in organic films is essential to discuss the performance of organic electronic devices such as organic light-emitting diodes (OLEDs) and organic photovoltaic cells (OPVs), which have undergone significant development in recent years.
Understanding elecronic structure at metal / organic and organic / organic interfaces is very important.
For example, the interface dipole and the band bending due to space charge have a significant impact on fundamental operating mechanisms such as carrier recombination and carrier generation. Furthermore, the interface charge induced by spontaneous orientation polarization (SOP)\cite{ito_spontaneous_2002, noguchi_spontaneous_2019, noguchi_understanding_2022, pakhomenko_understanding_2023} of the polar molecules is known to promote carrier injection\cite{noguchi_influence_2013} from the electrode and cause exciton quenching\cite{bangsund_subturn-exciton_2020, esaki_active_2021}. 
Investigation of the energy level alignment at the interface has been approached from both experimental and numerical aspects, and these methods have been advancing.

In terms of experimental techniques, photoelectron spectroscopy such as ultraviolet photoelectron spectroscopy (UPS), X-ray photoelectron spectroscopy (XPS), and Kelvin probe (KP) have so far been adopted to verify energy alignment at the interface\cite{ishii_energy_1999}. To determine the potential profile in a film, typically a film is deposited step by step and the surface potential is measured. In reality, however, the potential of the under layer also changes as the films are stacked. In other words, the sequence of surface potential values obtained by increasing film thickness in a step-wise manner do not recover the actual potential profile within the film\cite{arch_computer_1990,oehzelt_organic_2014}. Furthermore, in experiments, the substrate is often grounded, so the stacked films are placed in an external electric field-free situation, while in the actual device, an electric field corresponding to the work function difference between the anode and cathode is built in.
These are problems for every methods that probe the surface of films in general.

On the other hand, with respect to theoretical calculations, M. Oezelt et al. have proposed a method to simulate energy alignment at the electrode interface based on a thermal equilibrium model to achieves the determination of the potential within the film\cite{oehzelt_organic_2014, oehzelt_energy-level_2015}.
However, whether the thermal equilibrium model can be applied to organic materials remains controversial. In thermal equilibrium, the drift current and diffusion current are in balance, where Einstein's relation holds. Under such conditions, the Fermi level becomes constant regardless of location, and the Fermi-Dirac distribution holds throughout the entire system.
In organic semiconductors with a large band gap and rich trap states, only a limited number of thermally excited carriers are available in the bulk, and carrier mobility is prone to degradation owing to trap states.
As a result, the carrier distribution not necessarily reach thermal equilibrium within finite time, and the Fermi level often does not align with that of the electrode\cite{ishii_kelvin_2004}. Experimental verification of this problem is still limited.

Our group has been developing a rotary Kelvin probe (RKP)\cite{ohara_examination_2021, ohara_impact_2023}, which can measure the change in work function quasi-continuously with respect to the change in film thickness during deposition. RKP provides far more data points than conventional KP and allows for an accurate comparison of energy profile with simulations.

In this study, this comparison was made for a typical donor/acceptor system. We employed hexaazatriphenylene hexacarbonitrile (HAT-CN) as a hole injection layer (HIL) and N,N'-Di-1-naphthyl-N,N'-diphenylbenzidine (\aNPD) as a hole transport layer (HTL), where energy alignment has been actively investigated\cite{kim_energy_2009, lee_mechanism_2012, yang_role_2015, oh_energy_2017}. 
The two methods, experimental and simulation, resulted in significantly different profile of band bending in the \aNPD layer. This large difference originates from the limited and uneven distribution of thermally excited space charge, implying that Fermi level alignment across the film and electrode is not necessarily established.

\section{Experimental}
\subsection{Materials and RKP measurement}
The work function of the samples was measured with our custom made rotary Kelvin probe system; the RKP unit is installed in the deposition chamber (the base pressure $\leq 4\times 10^{-4}$ Pa), allowing simultaneous deposition onto the substrate and work function measurement.

Au substrates were ultrasonically cleaned twice with acetone and 2-propanol without UV ozone treatment.
The work function of Au substrate was determined with reference to a flexible graphite sheet (NeoGraf Grafoil\textregistered) whose work function was measured to be 4.55 eV by photoelectron yield spectroscopy (PYS).
HAT-CN and \aNPD (device grade) were generously provided by TOSOH Co., Ltd. and used without further purification. Both materials were vacuum-deposited at a deposition rate of 1 \AA/s. The sample was irradiated with UV light from a mercury xenon lamp (Hamamatsu L7212-01), whose emission spectrum extends below the absorption edge of HAT-CN\cite{liao_tandem_2008} and \aNPD\cite{yokoyama_horizontal_2008} (Figure \ref{UVspectrum}).
The film thickness measured by a quartz crystal microbalance was calibrated using a stylus step profiler (Kosaka Laboratory ET-4000A).

\subsection{Numerical calculation}
Numerical calculation of the electrostatic potential and space charge density was performed using a custom Python program based on the thermal equilibrium model proposed by M. Oezelt et al\cite{oehzelt_organic_2014, oehzelt_energy-level_2015}.
In their model, the potential distribution $V(z)$ is obtained by solving the Poisson equation
\begin{equation}
    \nabla[\varepsilon(z)\nabla V(z)]=-{\frac{\rho(z)}{\varepsilon_{0}}}
    \label{poisson}
\end{equation}
with $\varepsilon_{0}$ the vacuum permittivity and $\varepsilon$ relative permittivity.
Then the space charge density $\rho(z)$ is calculated from the density of states (DOS) distribution $D(E)$, fermi function $f(E)$ and $V(z)$

\begin{equation}
    \begin{split}
    \rho(z)=e\cdot n\cdot\left\{\int_{-\infty}^{+\infty}\mathrm{d}E\cdot f_{\mathrm{H}}(E)\cdot D_{\mathrm{H}}[E+e V(z)]\right.\\
    \left. - \int_{-\infty}^{+\infty}\mathrm{d}E\cdot f_{\mathrm{L}}(E)\cdot D_{\mathrm{L}}[E+e V(z)]\right\}
    \end{split}
    \label{rho}
\end{equation}

where, $e$ and $n$ represent the elementary charge and the number of molecules per unit volume, respectively.
The subscripts of $D$ and $f$ denote the highest occupied molecular orbital (HOMO) -derived and the lowest unoccupied molecular orbital (LUMO) -derived.
By such iterative updates of $V(z)$ and $\rho(z)$ , the energy level alignment is simulated self-consistently. 

The parameters of HAT-CN and \aNPD were determined from the literature as in Table \ref{param_table}. Some have reported the electron affinity of HAT-CN measured by inverse photoelectron spectroscopy to be above 4.9 eV\cite{yoshida_electron_2015} and others to be 5.42 eV\cite{lee_determination_2012}, while in the present study we used 5.2 eV, which is in good agreement with the experimental results.
The HOMO DOS and LUMO DOS of each material were assumed to be Gaussian functions with $\sigma = 0.25 \mathrm{eV}$.
In order to simulate the potential shift due to SOP of \aNPD, calculations were performed by placing \intcharge{+0.9} and \intcharge{-0.9} polarization charges on the surface of the \aNPD layer and at the \aNPD/HAT-CN layer interface, respectively.
This polarization charge generates a potential gradient of 3 mV/nm within the \aNPD layer.
All calculations were performed with the 1D mesh width of 1 nm.

\begin{table}
    \centering
    \begin{tabular}{cccccc}
        Layer & $E_{\mathrm{F}}$ (eV) & $\epsilon_{\mathrm{r}}$   & Site density ($\mathrm{cm}^{-3}$) & IE (eV) & EA (eV)  \\ \hline \hline
        Au & 4.64 & - &  - & - & - \\ \hline
        HAT-CN & - & 3.08  & $2.91\times 10^{21}$ & 9.31\cite{shimizu_study_2022} &  5.2  \\ \hline
        \aNPD & - &  3.35\cite{noguchi_charge_2012}  & $1.23\times 10^{21}$ & 5.18\cite{shimizu_study_2022}–5.78  & 1.9\cite{terashima_electronic_2020}  \\ \hline
    \end{tabular}
    \caption{Parameters used to calculate the electrostatic potential.}
    \label{param_table}
\end{table}

\section{Results and discussion}
\subsection{Energy level alignment of HAT-CN and \aNPD monolayer on Au substrate}
\begin{figure}[htbt]
    \centering
        \includegraphics[width=10cm]{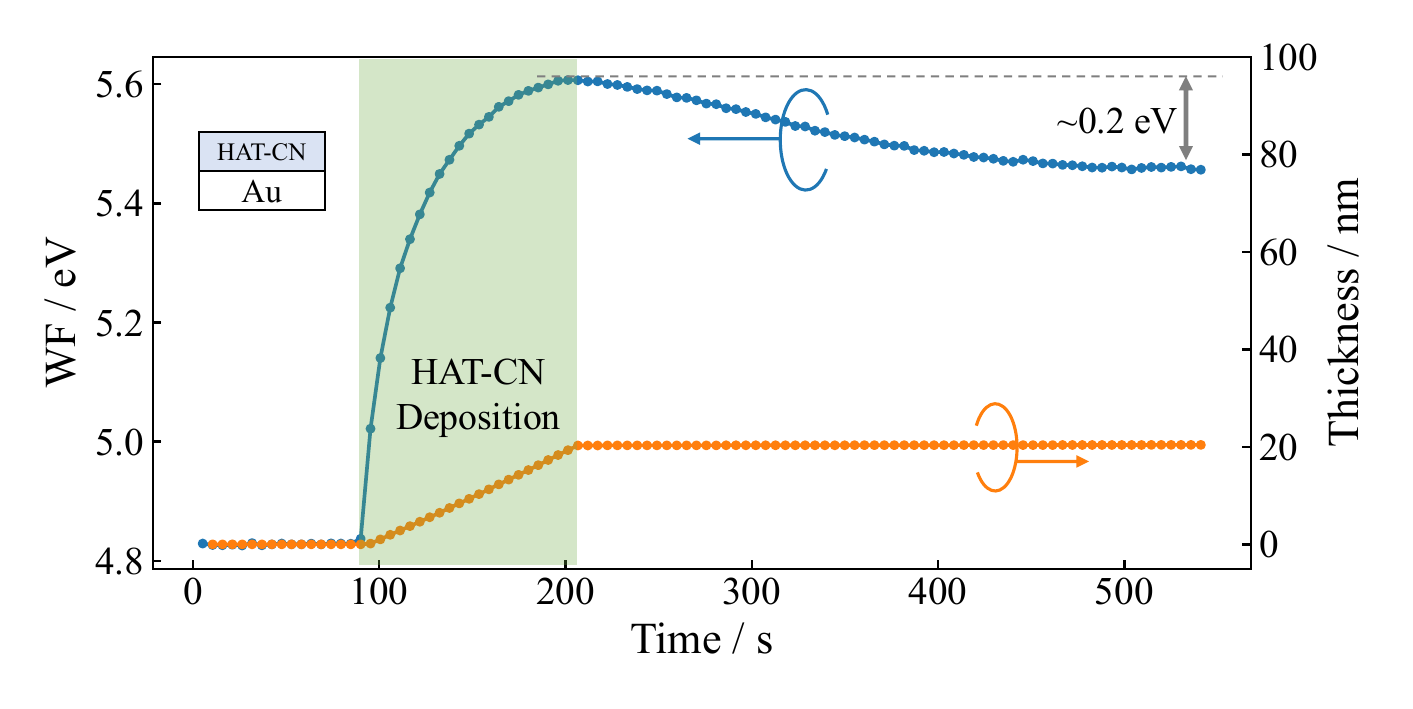}
        \caption{Work function (blue) and film thickness (orange) of HAT-CN films measured by the RKP system as a function of time.}
        \label{HAT-CN_woUV}
\end{figure}

Let's start with the results of HAT-CN deposition. Figure \ref{HAT-CN_woUV} shows the surface potential (blue) and the film thickness (orange) of HAT-CN deposited under dark conditions as a function of time.
The green-shaded region, where the film thickness increases, corresponds to the period during which deposition was performed. As the deposition progressed, the work function increased by approximately 0.8 eV. Since HAT-CN has a very deep LUMO level, it receives electrons from the Au substrate, forming a negative space charge layer near the interface with the substrate, which causes upward vacuum level shift. 

After the deposition was terminated, a gradual decrease in the work function was observed over a period of approximately 300 seconds. The maximum decay of the work function was about 0.2 eV. 
Possible mechanisms that explain this potential drop are (1) the return of charge from the HAT-CN film to the substrate and (2) the rearrangement of charges in the HAT-CN film. 
Low mobility carriers such as trapped electrons could return to the substrate over time. Otherwise, considering the capacitor model, this potential drop might be described by shifting the mass center of space charge toward the substrate.
Additionally, it has been reported that the molecular orientation of HAT-CN varies with film thickness\cite{joo_effects_2023}. The observed potential drop may also be attributable to changes in energy levels induced by temporal variations in molecular orientation.
In any case, such a system is also an example of a system that has not reached thermal equilibrium immediately after the deposition.

\begin{figure}[htbt]
    \centering
        \includegraphics[width=10cm]{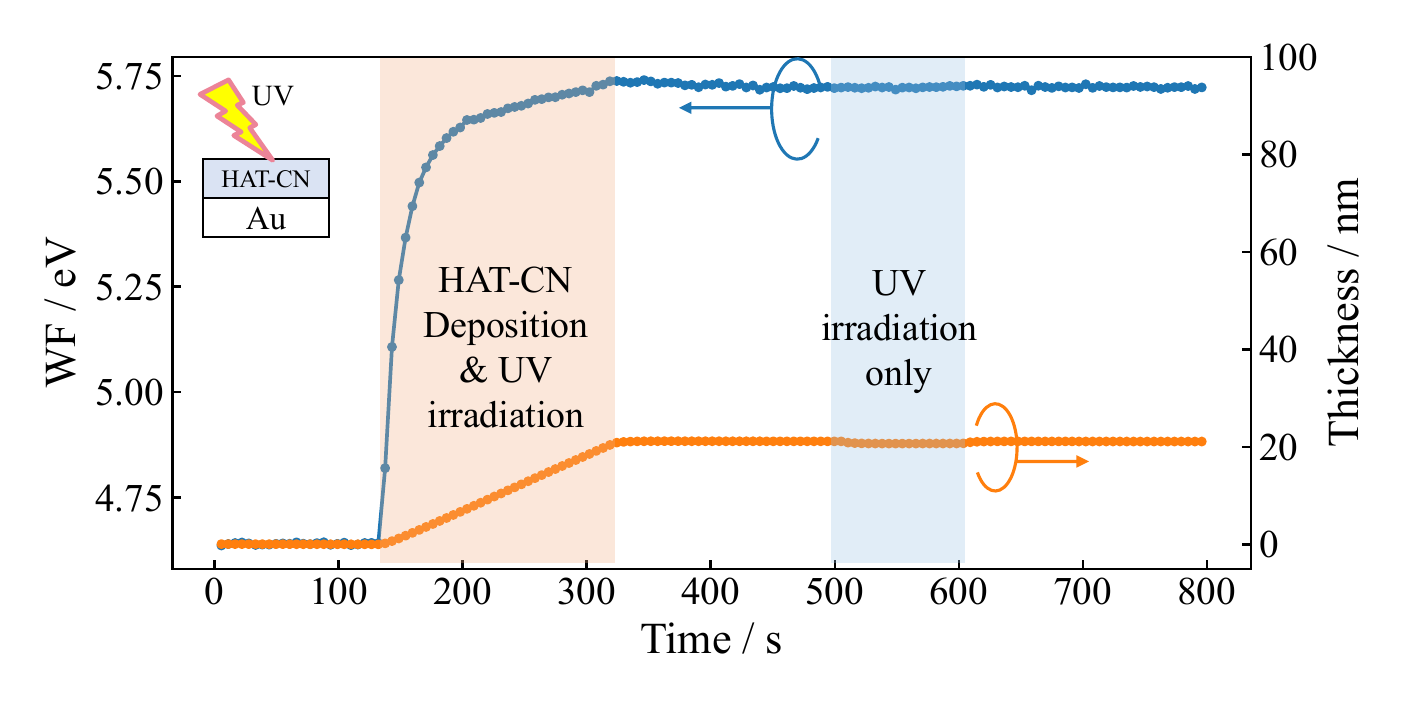}
        \caption{Time variation of work function (blue) and film thickness (orange) of HAT-CN films irradiated with UV during deposition.}
        \label{HAT-CN_wUV}
\end{figure}

The results of UV irradiation during deposition are then shown in Figure \ref{HAT-CN_wUV}. The orange region represents the time when deposition and UV irradiation were performed simultaneously, while the blue region represents the time when only UV irradiation was performed after deposition. The work function increased by approximately 1.1 eV as deposition progressed, and the decrease in work function after deposition was terminated was reduced to less than 30 mV. The net change in work function significantly differs from the 0.6 eV observed without UV irradiation, indicating that a greater amount of charge has been injected. This can be due to the charge rearrangement associated with carrier trapping/detrapping promoted by UV irradiation. Stimuli such as UV irradiation are helpful for the accurate measurement of energy level alignment, as they induce the system to transition into thermal equilibrium.

\begin{figure}[htbt]
    \centering
        \includegraphics[width=10cm]{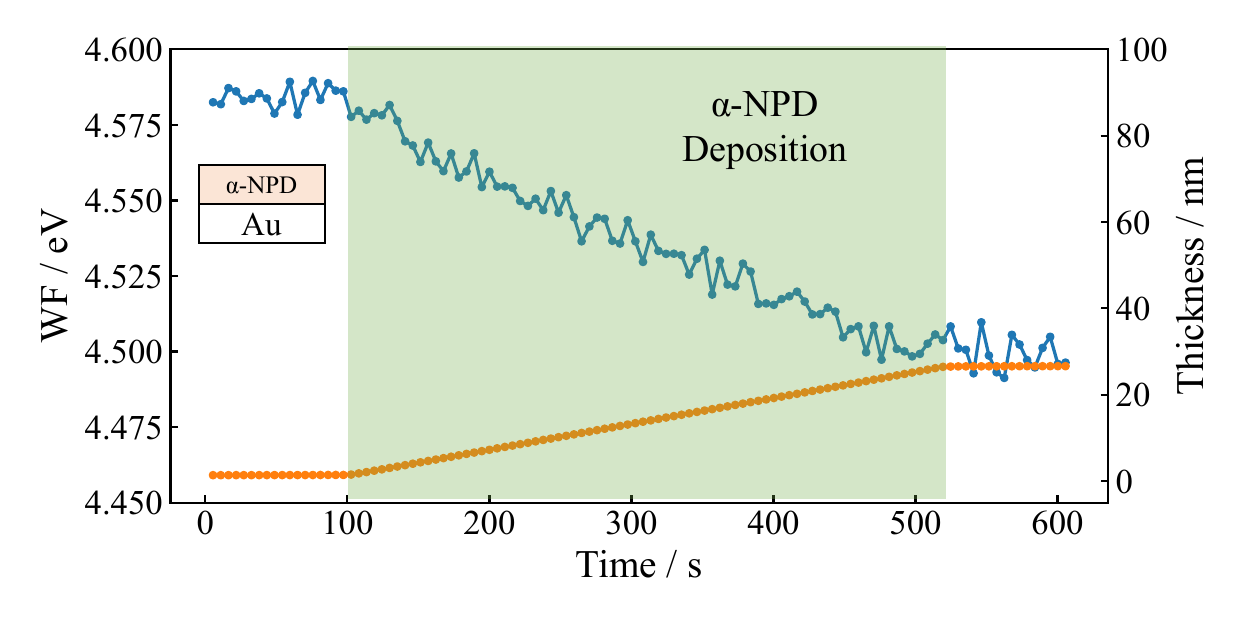}
        \caption{Time variation of work function (blue) and film thickness (orange) of \aNPD films.}
        \label{aNPD}
\end{figure}

The results of \aNPD deposition on the substrate are shown in Figure \ref{aNPD}. The work function changes linearly with increasing film thickness. This is recognized as being caused by the orientation polarization of \aNPD molecules, which have permanent dipole moment (PDM) of 0.5D. 
The resulting potential shift is referred to as giant surface potential (GSP), which has already been reported for \aNPD (\GSPslope{5.3}\cite{noguchi_charge_2012}).
From the observed potential change, the GSP slope was calculated to be \GSPslope{3.3}, which is in good agreement with previous studies: at the \aNPD/Au interface, band bending due to charge transfer hardly occurs and the potential changes only due to the orientation polarization.

\subsection{Energy level alignment of \aNPD/HAT-CN/Au multilayer film}
\begin{figure}[htbt]
    \centering
    \includegraphics[width=10cm]{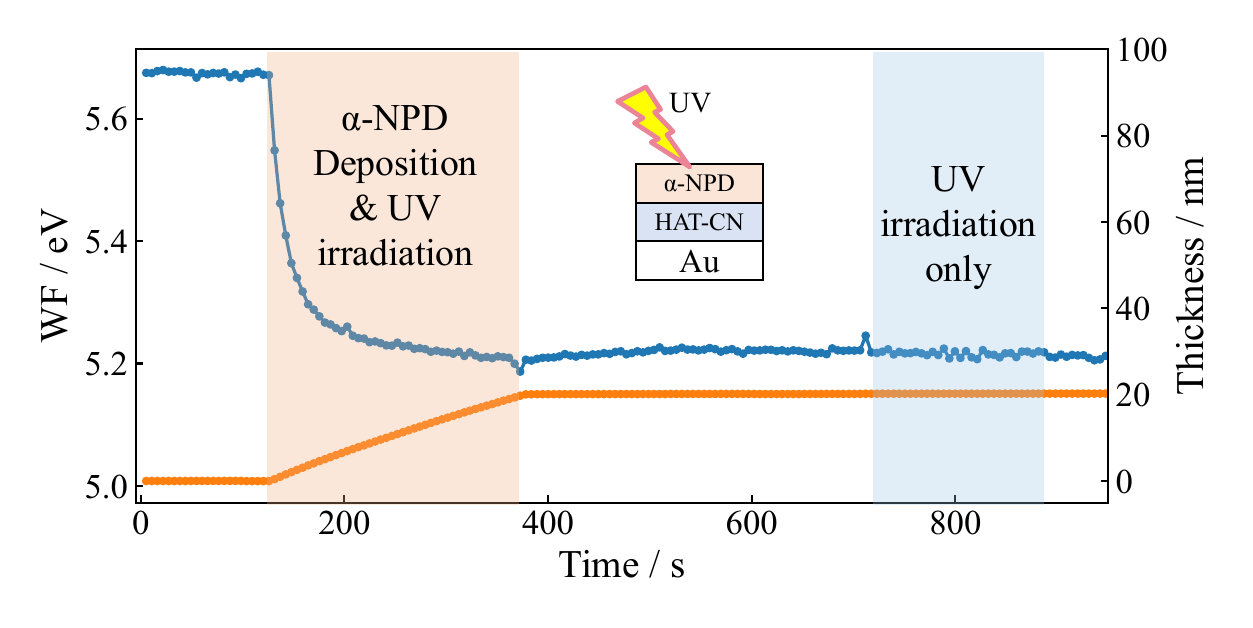}
    \caption{Time variation of work function (blue) and film thickness (orange) of \aNPD films on HAT-CN}
    \label{aNPDonHAT}
\end{figure}

Figure \ref{aNPDonHAT} shows the results of depositing \aNPD on HAT-CN samples which was brought close to thermal equilibrium through UV irradiation. 
During \aNPD film deposition, UV irradiation was performed simultaneously (orange region). After film deposition, only UV irradiation was performed (blue region).
Unlike the deposition on Au, it can be seen that the work function decreased abruptly immediately after the start of deposition, and then switched to a gentle change. It was confirmed that further UV irradiation does not induce any additional change in the work function.

\begin{figure}[htbt]
    \centering
    \includegraphics[width=11cm]{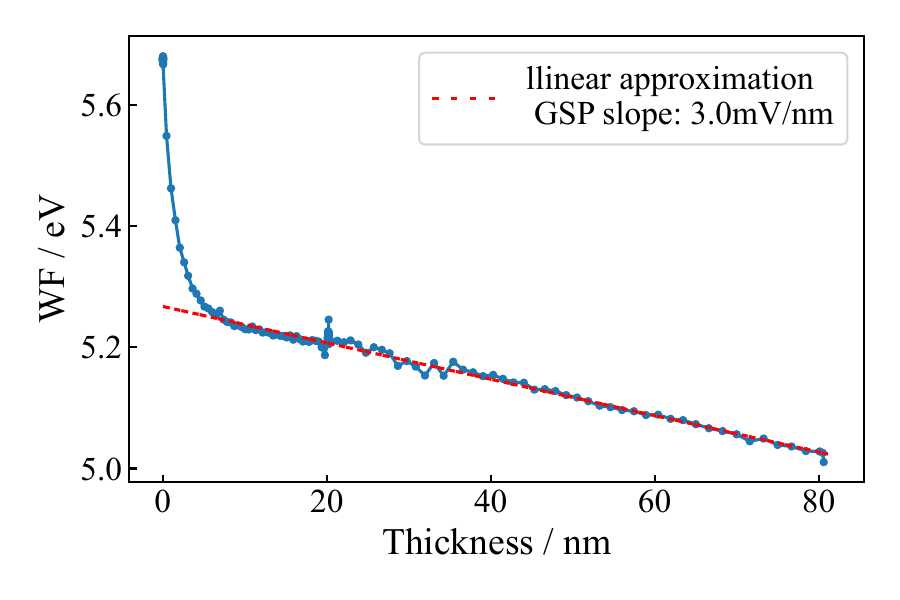}
    \caption{Work function within a 80 nm thick \aNPD film deposited on the UV irradiated HAT-CN film. }
    \label{aNPD_extend}
\end{figure}

As mentioned earlier, \aNPD film has a GSP, so the potential change caused by GSP and charge transfer needs to be treated separately. Therefore, the horizontal axis is redefined as film thickness, and the change in potential when \aNPD is deposited up to 60 nm is shown in Figure \ref{aNPD_extend}. The linear change seen in the range 10–20 nm continue up to 80 nm; the approximate line fitted to the linear function in the range 40–80 nm and extended to 0 nm is the red dashed line. The estimated GSP slope is \GSPslope{3.0}, which is in good agreement with the GSP magnitude of \aNPD on Au.

Notably, the deviation from the extrapolated straight line is limited to the range 0-6 nm. This suggests that there is almost no space charge in the bulk region. \aNPD molecules deposited on HAT-CN should theoretically receive holes to inject electrons to LUMO level of HAT-CN. However, the electric field induced by the SOP directs holes toward the \aNPD/HAT-CN interface rather than the surface, thereby preventing the transferred holes from spreading into the bulk region.
Therefore, at present it is possible to estimate a model in which the electric double layer is generated in the very close vicinity of the interface and the potential is shifted by the SOP from then onwards.

\begin{figure}[htbt]
    \centering
    \includegraphics[width=9.5cm]{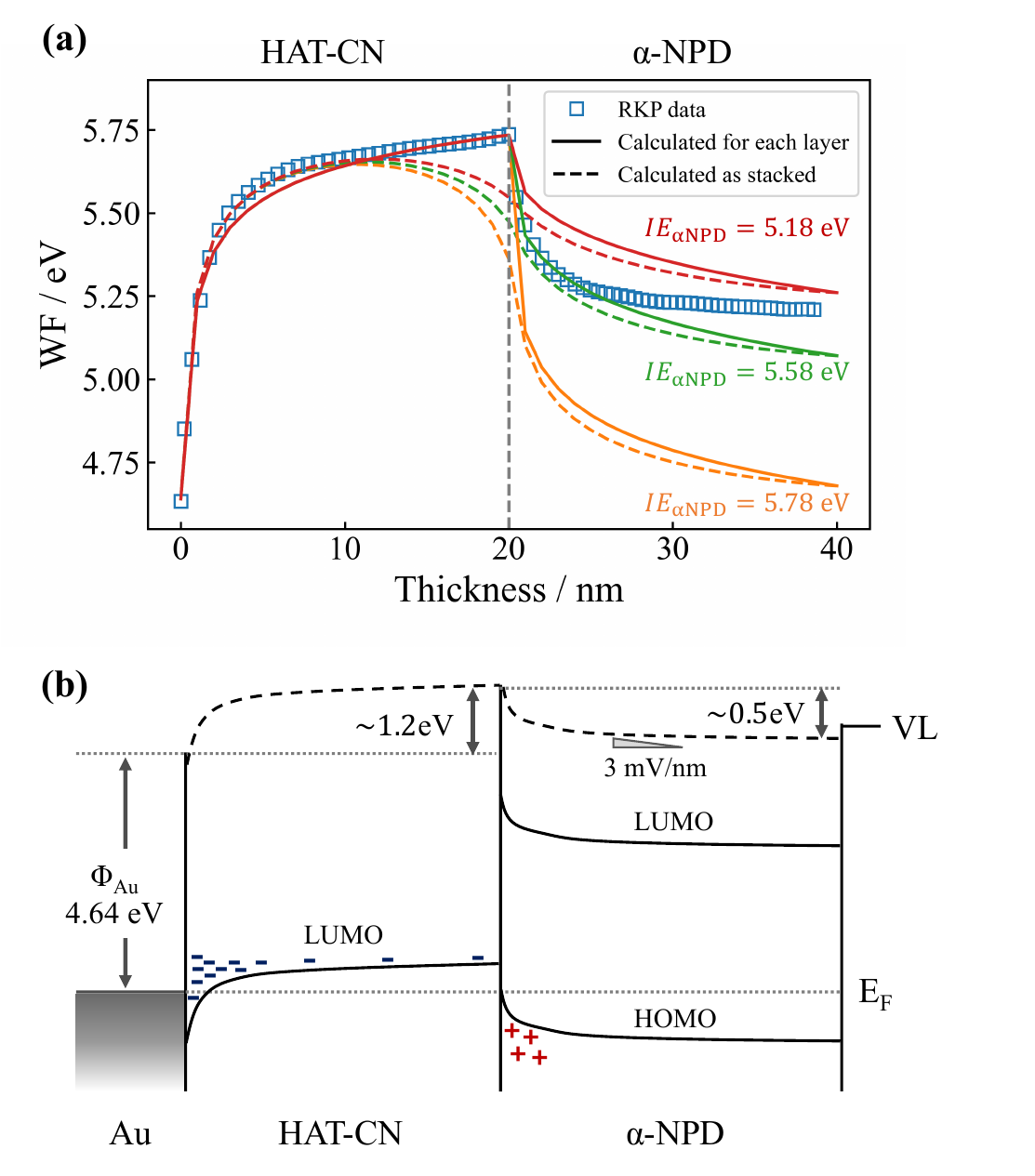}
    \caption{(a)Thickness dependence of the work function of the \aNPD/HAT-CN/Au interface and comparison with simulations. (b) A shematic illustration of energy diagram as inferred from the results of RKP measurements.}
    \label{HATNPD}
\end{figure}

Finally, the work function of the \aNPD/HAT-CN/Au interface as a function of film thickness and comparison with simulations is shown in Figure \ref{HATNPD}(a). The blue squares show the work functions measured by our RKP. The orange, green and red solid lines show the work functions obtained by calculating the surface potentials while increasing the thickness of the organic films in steps, corresponding to IE of 5.18 eV, 5.58 eV and 5.78 eV for \aNPD, respectively. The dashed lines represent calculated potential within the films. Figure \ref{HATNPD}(b) represents a shematic illustration of energy diagram as inferred from the results of RKP measurements.

If we start from the HAT-CN layer, it can be seen that the simulations and measurements are in good agreement. HAT-CN possesses a LUMO level deeper than the work function of the Au substrate, and the enhanced charge transfer induced by UV irradiation is presumed to bring the system close to thermal equilibrium. It should be noted, however, that thermal equilibrium on a metal substrate is not always necessarily achieved. For example, the GSP induced within the film by SOP generally increases in proportion to the film thickness and can, in some cases, cause potential shifts of several tens of electronvolts\cite{ito_spontaneous_2002}. In such a situation, not only the LUMO level but even the vacuum level would lie far below the Fermi level of the substrate. If the Fermi level were assumed to be aligned throughout the film, an astronomically large number of thermally excited carriers would be induced, which is, however, physically unrealistic. Systems in which GSP is observed represent typical non-equilibrium cases, and the deviation from the thermal equilibrium model has already been examined\cite{wang_stabilization_2023}.

There are two possible reasons for the slightly upward position of the measured values in the range 5–10 nm.
The first is that the space charge layer in the HAT-CN is localised relatively close to the interface with the substrate. The charge distribution may change as low mobility carriers such as trapped electrons move towards the interface.
The second case concerns the influence of simulation parameters. For example, the shape of the HOMO and LUMO DOS is assumed to follow a single Gaussian distribution in this study. In reality, however, due to molecular packing disorder and the presence of impurities, low density of states form within the bandgap\cite{sueyoshi_low-density_2009,yang_origin_2017}. Although the intensity of these in-gap states is low, their proximity to the Fermi level can lead to the generation of a non-negligible number of thermally excited carriers, which may in turn affect band bending.
Therefore, it is necessary to take into account the in-gap states of HAT-CN when determining the DOS shape. For a more accurate evaluation, it would be helpful to incorporate the experimentally measured occupied/unoccupied DOS obtained by photoelectron spectroscopic techniques. However, the trap-like nature of such in-gap states suggests that the Femi level alignment is still not fully established. 

Turning to the \aNPD layer, the thermal equilibrium model do not reproduce the measurements for any IE of \aNPD. The difference is particularly pronounced in the bulk region. Althogh, in the simulation, carriers can be thermally excited even within the bulk of \aNPD layer, in practice they are supplied exclusively by HAT-CN layer resulting charge transfer is extremely limited.
As mentioned before, The sample is placed under an external electric field-free environment and there are no forces actively pushing the carriers into the bulk of the \aNPD layer. Rather, due to SOP of \aNPD, negative and positive polarization charges are induced in the HAT-CN/\aNPD interface and surface, respectively. These charges could attract and trap holes at the HAT-CN/\aNPD interface, preventing their diffusion into the bulk. In such case, it is necessary to promote carrier injection not only by UV irradiation but also by applying an external electric field. 

These findings demonstrate that the assumption of a single Fermi level across the organic film is not valid, and energy level alignment is not achieved through thermally excited carriers, but rather determined by how carriers—supplied exclusively from the substrate—propagate within the film.

\section{Conclusions}
In the present work, to address the issues of conventional theoretical and experimental investigations in energy level alignment models for metal/organic and organic/organic interfaces, the work function of donner/acceptor system was measured quasi-continuously with respect to the film thickness by using RKP and the results were compared with simulations assuming a thermal equilibrium model.

For the metal/organic HAT-CN/Au interface, the experimental results were consistent with the simulations only when UV irradiation was applied during deposition, demonstrating that external stimulation is often required to promote charge transfer/transport and bring the system closer to thermal equilibrium. In contrast, for the organic/organic \aNPD/HAT-CN interface, the experimental results deviated significantly from the model predictions. The potential changes were dominated by SOP of \aNPD, and charge transfer into the bulk region was strongly suppressed. These observations indicate that the carriers responsible for energy level alignment are supplied primarily from the substrate, rather than being generated thermally within the organic bulk, and their propagation is strongly influenced by interfacial polarization effects. Consequently, the conventional assumption of Fermi level alignment throughout organic multilayers is not practically valid. These findings highlight the limitations of applying a simple thermal equilibrium model to organic heterointerfaces and emphasize the importance of considering nonequilibrium effects due to traps, in-gap states, and external stimuli such as UV irradiation or applied electric fields. 

The present results demonstrate the utility of RKP as a powerful tool for probing interfacial energetics in organic films and provide new insights into the mechanisms that determine energy level alignment in donor/acceptor systems, which will be valuable for the rational design of organic electronic devices with improved charge injection and transport characteristics.

\section*{Acknowledgement}
We thank TOSOH Co., Ltd. for providing the HAT-CN and \aNPD molecules. This research was supported by a Japan Society for the Promotion of Science (JSPS) KAKENHI Grant-in-Aid for Scientific Research (grant no. 24K0156) and Grant-in-Aid for JSPS Fellows (grant no. 22J21883). 

\appendix

\section{Supplementary data}
\begin{figure}[htbt]
    \centering
    \includegraphics[width=11cm]{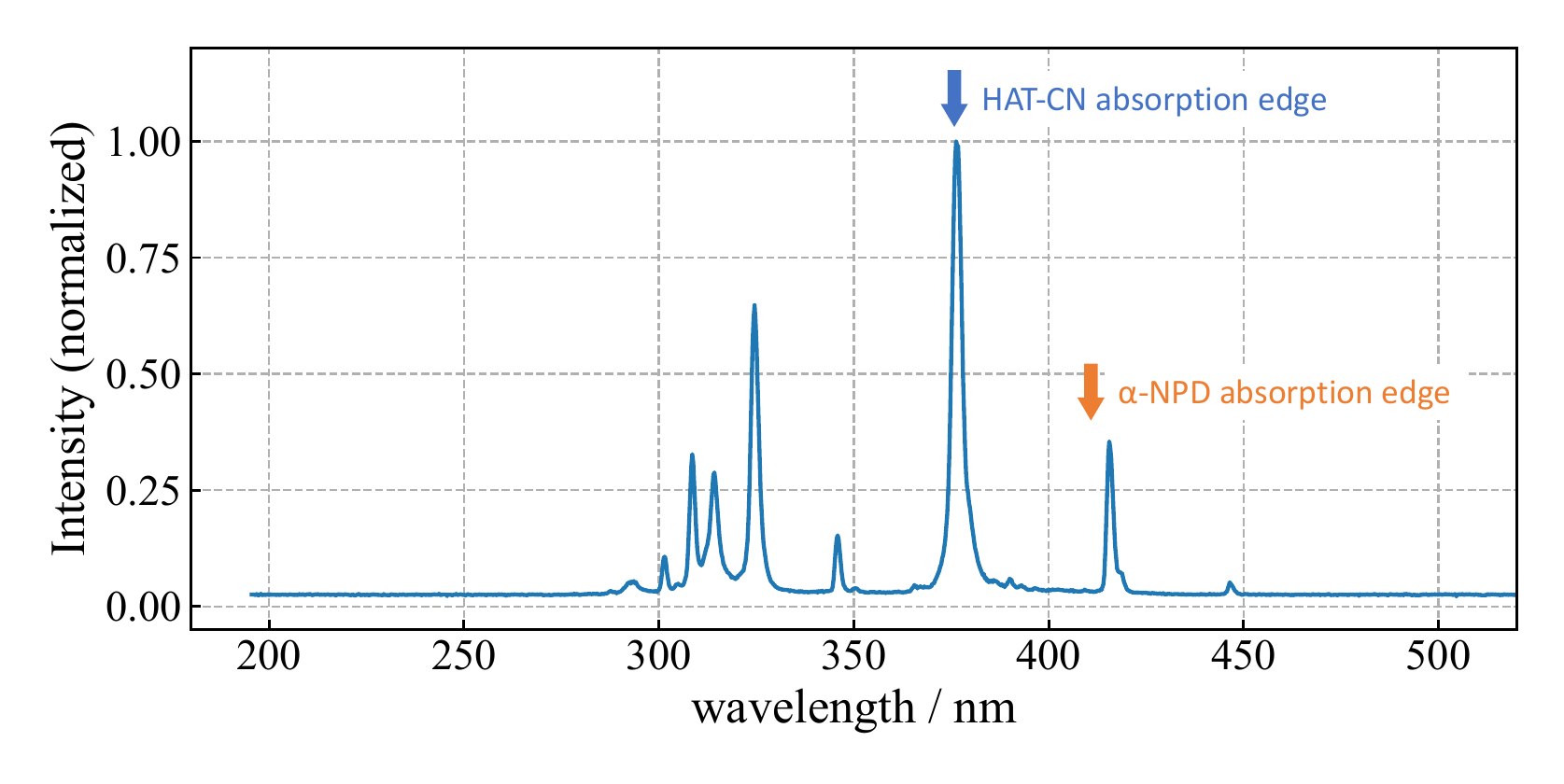}
    \caption{Measured emission spectrum of the mercury xenon lamp (Hamamatsu L7212-01)}
    \label{UVspectrum}
\end{figure}

\bibliographystyle{elsarticle-num} 
\bibliography{Mybib}

\end{document}